\documentclass[final,12pt,sort&compress]{elsarticle}
\usepackage[T1]{fontenc}
\usepackage[latin9]{inputenc}
\usepackage{amsmath}
\usepackage{amssymb}
\usepackage{graphicx}

\makeatletter
\journal{Physica B: Condensed Matter}

\makeatother

\begin{document}

\begin{frontmatter}{}

\title{Quantum gap and spin-wave excitations\\
in the Kitaev model on a triangular lattice}

\author[uni,cnrsa,cnism]{Adolfo Avella}

\author[uni]{Andrea Di Ciolo}

\author[Stutt-uni,MPI]{George Jackeli}

\address[uni]{Dipartimento di Fisica ``E.R. Caianiello'',\\
Universit\`a degli Studi di Salerno, I-84084 Fisciano (SA), Italy}

\address[cnrsa]{CNR-SPIN, UOS di Salerno, I-84084 Fisciano (SA), Italy}

\address[cnism]{Unit\`a CNISM di Salerno, Universit\`a degli Studi di Salerno,
I-84084 Fisciano (SA), Italy}

\address[Stutt-uni]{ Institute for Functional Matter and Quantum Technologies,\\
University of Stuttgart, Pfaffenwaldring 57, D-70569 Stuttgart, Germany}

\address[MPI]{Max Planck Institute for Solid State Research,\\
Heisenbergstrasse 1, D-70569 Stuttgart, Germany}
\begin{abstract}
We study the effects of quantum fluctuations on the dynamical generation
of a gap and on the evolution of the spin-wave spectra of a frustrated
magnet on a triangular lattice with bond-dependent Ising couplings,
analog of the Kitaev honeycomb model. The quantum fluctuations lift
the subextensive degeneracy of the classical ground-state manifold
by a quantum order-by-disorder mechanism. Nearest-neighbor chains
remain decoupled and the surviving discrete degeneracy of the ground
state is protected by a hidden model symmetry. We show how the four-spin
interaction, emergent from the fluctuations, generates a spin gap
shifting the nodal lines of the linear spin-wave spectrum to finite
energies.
\end{abstract}

\end{frontmatter}{}

\section{Introduction}

Accidental degeneracies between various order patterns are a characteristic
of frustrated magnets, where not all pairwise exchange interactions
can be simultaneously satisfied \citep{Bal10}. Order-by-disorder
mechanism, driven by thermal and/or quantum fluctuations, is often
capable to lift such degeneracies selecting an unique ground state
\citep{Vil80,She82,Sav12}. Actually, the order-by-disorder mechanism
results inactive on highly frustrated quantum magnets (e.g., Kagomé
and pyrochlore isotropic spin one-half antiferromagnets) and these
latter remain disordered down to the lowest temperatures, realizing
the so-called quantum spin liquid in their ground states \citep{Bal10}.
Another relevant case in which it is possible to have a spin-liquid
ground state is Mott insulators with unquenched orbital moments and
strong spin-orbit coupling where bond-dependent Ising-type interactions
may dominate over the conventional Heisenberg term and, even being
ferromagnetic, can frustrate long-range magnetic orders \citep{Jac09}.
The exactly solvable Kitaev honeycomb model \citep{Kit06}, where
nearest-neighbor spins are coupled by Ising-type terms selected by
the bond directionality (each of the three spin components for each
of the three non-equivalent bonds in the lattice), is the most famous
theoretical realization of this scenario.

Several extensions of the Kitaev model have been studied and, in particular,
the so-called Kitaev-Heisenberg model \citep{Cha10,Jia11,Reu11,Pri12,Cha13}
in connection to many experimental facts \citep{Sin10,Liu11,Cho12,Sin12,Ye12,Gre13}.
The resulting theoretical phase diagram is very rich and includes
both the ordered phases seen experimentally and the quantum spin-liquid
around the Kitaev limit. Recently, a triangular analog of the Kitaev-Heisenberg
model \textendash{} an extension of the anisotropic spin model originally
proposed to study sodium cobaltates \citep{Kha05} \textendash{} for
classical \citep{Rou16} and quantum \citep{Bec15,Li15} spins has
been studied numerically. The obtained rich phase diagram includes
a nematic phase of decoupled Ising chains with sub-extensive degeneracy
at the Kitaev limit \citep{Bec15}. As the true nature of the ground
state for quantum spins in the Kitaev triangular model, despite the
intensive numerical analyses, was still not understood, two of the
authors studied such model and solved the puzzle of its ground state
\textendash{} explaining and quantifying the results obtained by previous
numerical simulations \textendash{} by analyzing the effects of quantum
fluctuations within both the linked-cluster expansion \citep{Gel00},
combined with degenerate perturbation theory, and the linear spin-wave
theory \citep{Jackeli_15}. In particular, they have shown (a) the
presence of a mechanism of quantum selection of an easy axis that
reduces the classical degeneracy, (b) the emergence of a next-nearest-neighbor
four-spin interaction that reduces the sub-extensive degeneracy of
the ground state manifold, and (c) the existence of a hidden symmetry
of the model that protects the remaining degree of degeneracy.

In this short paper, we study the effects of the quantum fluctuations
on the dynamical generation of a quantum gap and on the evolution
of the spin-wave spectrum driven by the next-nearest-neighbor four-spin
interaction found previously \citep{Jackeli_15}. Such a coupling,
emergent from the quantum corrections to the classical ground state,
effectively induces a quantum spin gap shifting the nodal lines of
the previously found linear spin-wave spectrum to finite energies.

\begin{figure}
\noindent \begin{centering}
\includegraphics[height=4.5cm]{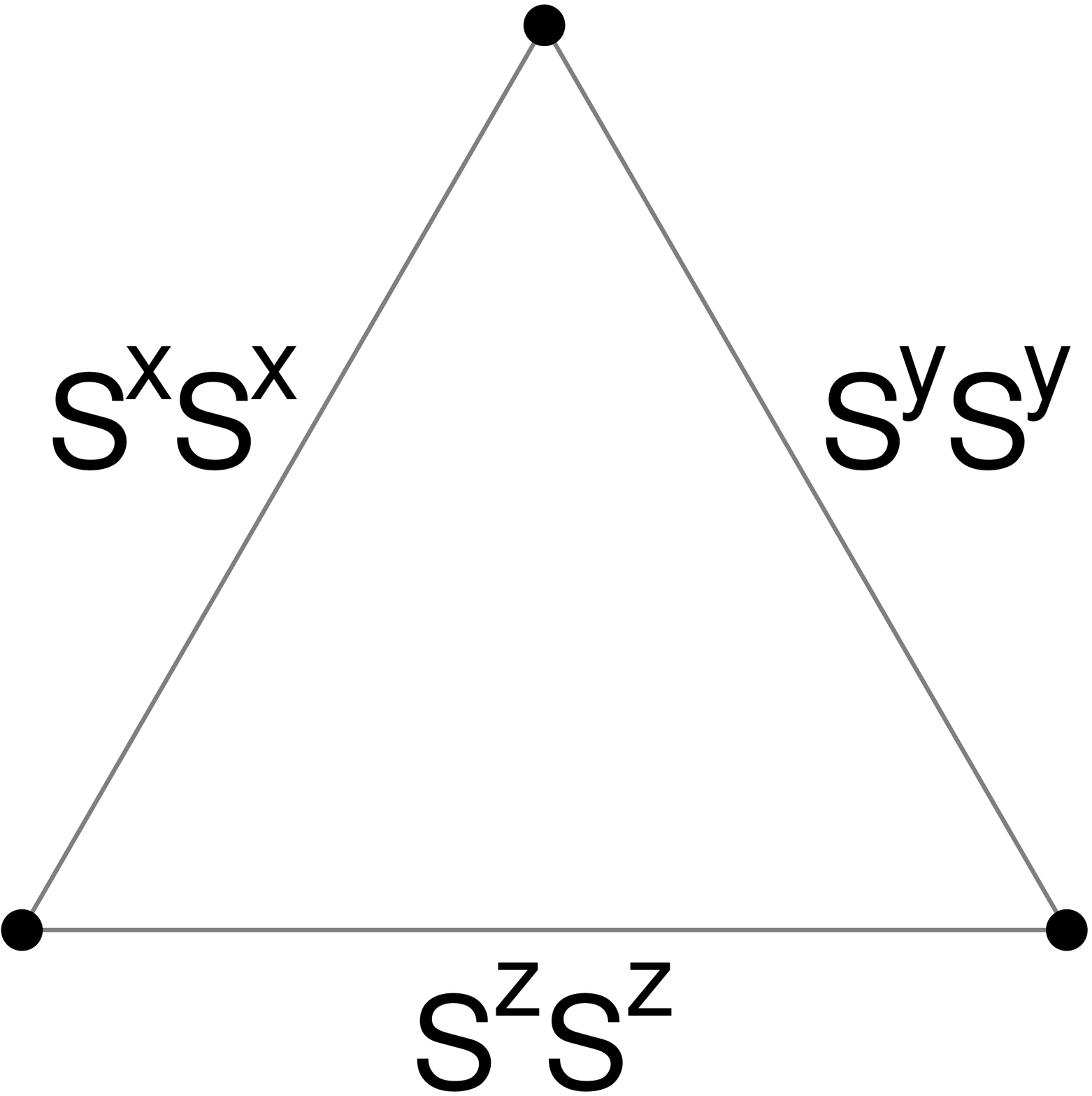}\hspace*{\fill}\includegraphics[height=4.5cm]{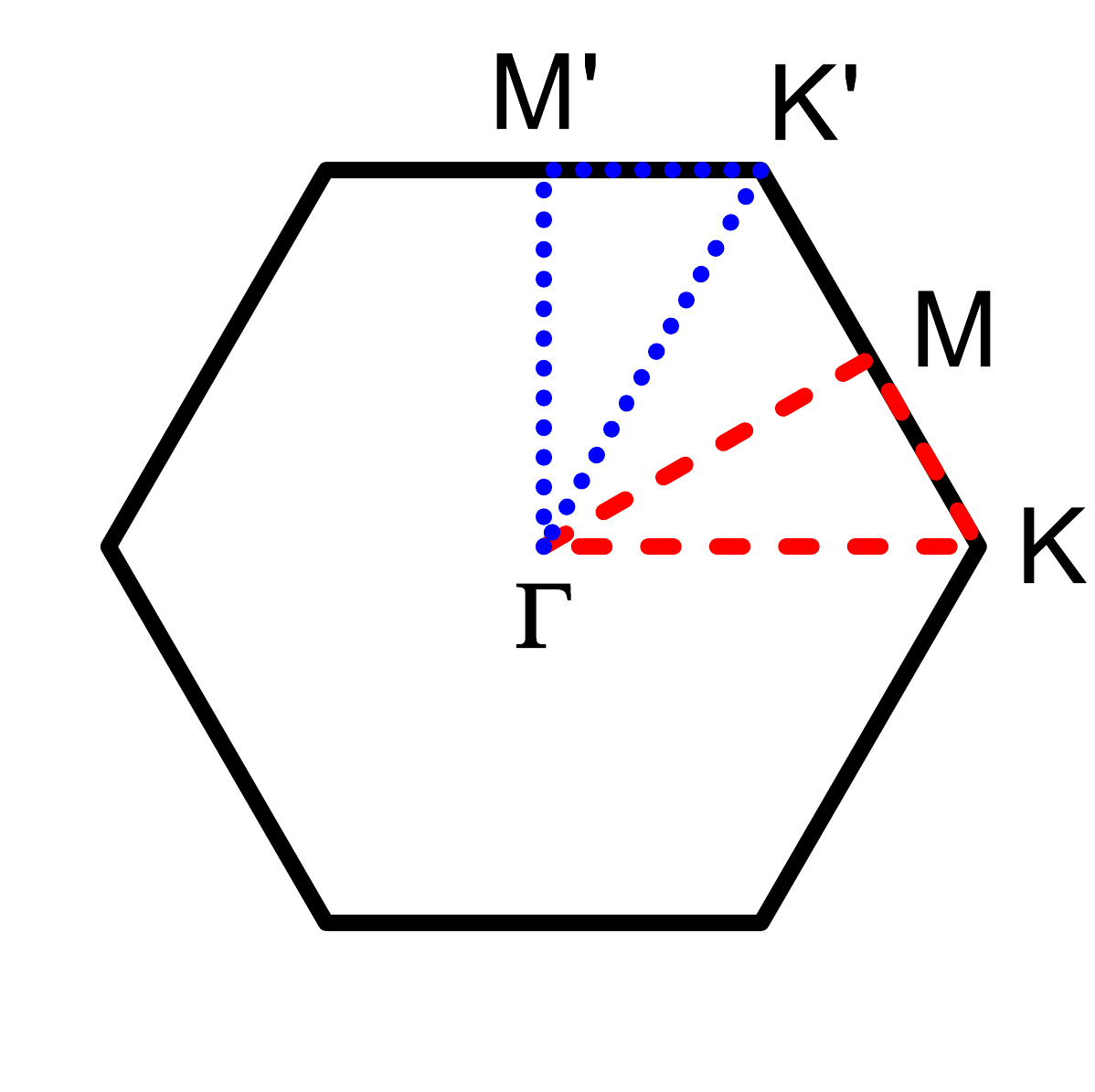}
\par\end{centering}
\caption{(left) Unit cell of the triangular lattice of model (\ref{Ham}) where
the Ising-type nearest-neighbor spin couplings on the three non-equivalent
bonds \textendash{} $(\gamma)$-bonds are each perpendicular to the
$\gamma\:(=x,\:y,\:z)$ spin quantization axis \textendash{} have
only the $\gamma$-component. (right) First Brillouin zone of the
same triangular lattice and principal directions used to represent
the spin-wave spectrum: $\Gamma=\left(0,0\right)$, $M=\left(\pi,\pi/\sqrt{3}\right)$,
$K=\left(4\pi/3,0\right)$, $M'=\left(0,2\pi/\sqrt{3}\right)$, $K'=\left(2\pi/3,2\pi/\sqrt{3}\right)$.\label{fig1}}
\end{figure}

\begin{figure}
\noindent \begin{centering}
\includegraphics[width=1\columnwidth]{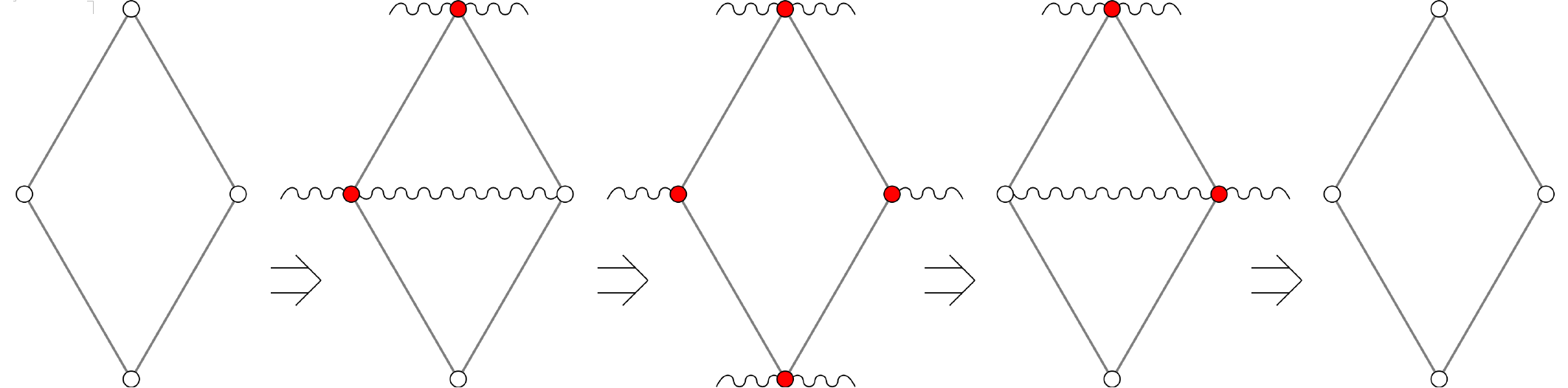}
\par\end{centering}
\caption{Schematic view of the $4^{th}$ order perturbation process leading
to the coupling of the four spins sitting around a diamond, by means
of quantum fluctuations {[}see Eq.~\eqref{pert}{]}. At each step,
spin flips are performed in pairs at one of the diamond four bonds:
in the virtual states the location of the misaligned spins are indicated
by filled circles and the broken ($z$)-bonds are marked by wavy lines.\label{fig2}}
\end{figure}

\section{Model }

We consider a spin one-half system (we keep $S$ generic for notational
convenience) residing on a triangular lattice lying in the $(1,1,1)$
plane of the spin-quantization frame {[}see Fig.~\ref{fig1} (left){]}.
The label $(\gamma)\:(=x,\:y,\:z)$ refers to its three non-equivalent
nearest-neighbor bonds spanned by the lattice vectors $\mathbf{a}_{x}=\left(1/2,-\sqrt{3}/2\right)$,
$\mathbf{a}_{y}=\left(1/2,\sqrt{3}/2\right)$ and $\mathbf{a}_{z}=\left(1,0\right)$,
respectively {[}see Fig.~\ref{fig1} (left){]}. The $(\gamma)$-bond,
which is perpendicular to the $\gamma$ spin-quantization axis, hosts
nearest-neighbor spin couplings only between the $S_{\mathbf{i}}^{\gamma}$
components of the spin operators $\mathbf{S}_{\mathbf{i}}$ {[}see
Fig.~\ref{fig1}(left){]}, and the corresponding Hamiltonian takes
the following form \citep{Jackeli_15}
\begin{equation}
H=-\sum_{\mathbf{i},\gamma}K_{\gamma}S_{\mathbf{i}}^{\gamma}S_{\mathbf{i}+\mathbf{a}_{\gamma}}^{\gamma}\,.\label{Ham}
\end{equation}
Given that the signs of the $K_{\gamma}$ couplings can be individually
flipped by means of a canonical transformation \citep{Jackeli_15},
without any loss of generality, we consider hereafter all $K_{\gamma}$
to be positive (ferromagnetic couplings). Then, after the analysis
performed in Ref.~\citep{Jackeli_15} within the linked-cluster expansion
\citep{Gel00}, we know that a coupling between pairs of spins belonging
to next-nearest-neighbor chains emerges by the quantum corrections
induced by the $K_{x}$ and $K_{y}$ terms to the ferromagnetic classical
ground state at the $4^{th}$ order in a \emph{diamond}-shape 4-site
cluster (see Fig.~\ref{fig2}). The so derived coupling has the following
expression
\begin{equation}
\delta H=-K\sum_{\mathbf{i}}\left(S_{\mathbf{i}}^{z}S_{\mathbf{i}+\mathbf{a}_{z}}^{z}\right)S_{\mathbf{i}+\mathbf{a}_{x}}^{z}S_{\mathbf{i}+\mathbf{a}_{y}}^{z}\label{pert}
\end{equation}
where $K=\frac{1}{24}\frac{K_{x}^{2}K_{y}^{2}}{\left|K_{z}^{3}\right|}$
and the sites $\mathbf{i}+\mathbf{a}_{x}$ and $\mathbf{\mathbf{i}}+\mathbf{\mathbf{a}}_{y}$
belong to next-nearest-neighbor chains {[}they are actually the ends
of the longer diagonal of the \emph{diamond} cluster, see Fig.~\ref{fig2}{]}.
Given that $S_{\mathbf{i}}^{z}S_{\mathbf{i}+\mathbf{a}_{z}}^{z}$
is just $\frac{1}{4}$ for $K_{z}>0$, the coupling between next-nearest-neighbor
chains and the one acting along the chains have the same \emph{sign}.
Taking into account such corrections, we come to the effective Hamiltonian
we wish to analyze in this short paper within the linear spin-wave
theory:
\begin{equation}
H'=H+\delta H.\label{HamT}
\end{equation}
It is worth noting that, within the linear spin-wave theory, $\delta H$
provides higher-order terms with respect to those emerging from $H$.
Accordingly, our treatment of Eq.~\eqref{HamT} automatically and
exactly avoids any double counting although, obviously, an exact treatment
of $H$ will give also the contribution coming from $\delta H$.

\section{Spin-wave theory}

\subsection{\textmd{\normalsize{}Harmonic approximation}}

First, we apply the linear spin-wave theory to the Hamiltonian \eqref{Ham},
\textit{i.e. }keeping only terms bilinear in the $a$-operators and
thus obtaining an $\mathcal{O}\left(1/S\right)$ expansion \citep{Jackeli_15}.
In particular, we consider the ferromagnetic state and use the Holstein-Primakoff
(HP) transformation
\begin{align}
 & S_{\mathbf{i}}^{z}=S-a_{\mathbf{i}}^{\dagger}a_{\mathbf{i}}\\
 & S_{\mathbf{i}}^{x}=\sqrt{\dfrac{S}{2}}\left(a_{\mathbf{i}}+a_{\mathbf{i}}^{\dagger}\right)\\
 & S_{\mathbf{i}}^{y}=\dfrac{1}{i}\sqrt{\dfrac{S}{2}}\left(a_{\mathbf{i}}-a_{\mathbf{i}}^{\dagger}\right)
\end{align}
where the bosonic $a_{\mathbf{i}}$ operators, sited at the site $\mathbf{i}$
of the lattice, satisfy canonical commutation relations $\left[a_{\mathbf{i}},a_{\mathbf{j}}^{\dagger}\right]=\delta_{\mathbf{ij}}$
and $\left[a_{\mathbf{i}},a_{\mathbf{j}}\right]=\left[a_{\mathbf{i}}^{\dagger},a_{\mathbf{j}}^{\dagger}\right]=0$.
In this representation, the Hamiltonian \eqref{Ham} reads as
\begin{align}
H & =-K_{x}\dfrac{S}{2}\sum_{\mathbf{i}}\left(a_{\mathbf{i}}^{\dagger}a_{\mathbf{i}+\mathbf{a}_{x}}+a_{\mathbf{i+}\mathbf{a}_{x}}^{\dagger}a_{\mathbf{i}}+a_{\mathbf{i}}^{\dagger}a_{\mathbf{i+}\mathbf{a}_{x}}^{\dagger}+a_{\mathbf{i}}a_{\mathbf{i}+\mathbf{a}_{x}}\right)\nonumber \\
 & -K_{y}\dfrac{S}{2}\sum_{\mathbf{i}}\left(a_{\mathbf{i}}^{\dagger}a_{\mathbf{i}+\mathbf{a}_{y}}+a_{\mathbf{i+}\mathbf{a}_{y}}^{\dagger}a_{\mathbf{i}}-a_{\mathbf{i}}^{\dagger}a_{\mathbf{i+}\mathbf{a}_{y}}^{\dagger}-a_{\mathbf{i}}a_{\mathbf{i}+\mathbf{a}_{y}}\right)\nonumber \\
 & -NK_{z}S^{2}+K_{z}S\sum_{\mathbf{i}}\left(a_{\mathbf{i}}^{\dagger}a_{\mathbf{i}}+a_{\mathbf{i+}\mathbf{a}_{z}}^{\dagger}a_{\mathbf{i}+\mathbf{a}_{z}}\right)
\end{align}
where $N$ is the number of the lattice sites. Then, using the Fourier
transform $a_{\mathbf{i}}=\dfrac{1}{\sqrt{N}}\sum_{\mathbf{k}}e^{i\mathbf{k}\cdot\mathbf{r_{i}}}a_{\mathbf{k}}$
and, therefore, moving to the momentum space, we find 
\begin{equation}
H=-NK_{z}S^{2}+\sum_{\mathbf{k}}\left[A_{0}\left(\mathbf{k}\right)a_{\mathbf{k}}^{\dagger}a_{\mathbf{k}}-\dfrac{1}{2}B_{0}\left(\mathbf{k}\right)\left(a_{\mathbf{k}}^{\dagger}a_{-\mathbf{k}}^{\dagger}+a_{\mathbf{k}}a_{-\mathbf{k}}\right)\right],
\end{equation}
 where $A_{0}\left(\mathbf{k}\right)=S\left(2K_{z}-K_{x}\cos\mathbf{k}_{\mathbf{a}_{x}}-K_{y}\cos\mathbf{k}_{\mathbf{a}_{y}}\right)$,
$B_{0}\left(\mathbf{k}\right)=S\left(K_{x}\cos\mathbf{k}_{\mathbf{a}_{x}}-K_{y}\cos\mathbf{k}_{\mathbf{a}_{y}}\right)$,
$\mathbf{k}_{\mathbf{a}_{x}}=\mathbf{k}\cdot\mathbf{a}_{x}$ and $\mathbf{k}_{\mathbf{a}_{y}}=\mathbf{k}\cdot\mathbf{a}_{y}$.
Accordingly, the magnonic spectrum $\omega\left(\mathbf{k}\right)$
is given by 
\begin{align}
\omega\left(\mathbf{k}\right) & =\sqrt{A_{0}^{2}\left(\mathbf{k}\right)-B_{0}^{2}\left(\mathbf{k}\right)}=2S\sqrt{\left(K_{z}-K_{x}\cos\mathbf{k}_{\mathbf{a}_{x}}\right)\left(K_{z}-K_{y}\cos\mathbf{k}_{\mathbf{a}_{y}}\right)}.\label{wk0}
\end{align}
It is worth reminding that the ferromagnetic state results the lowest-energy
one once introducing the quantum corrections on top of the classical
ground state \citep{Jackeli_15}.

\subsection{\textmd{\normalsize{}Magnon interactions}}

Then, we intend to see how the spin-wave dispersion is modified upon
introducing the higher-order term $\delta H$ \eqref{pert}, leading
to the effective Hamiltonian $H'$. To this aim, we apply the linear
spin-wave theory to this latter \eqref{HamT} and consider again the
ferromagnetic state. Keeping only terms bilinear in the $a$-operators,
we obtain the following expression for the higher-order term $\delta H$
in real space 
\begin{equation}
\delta H=-NKS^{4}+KS^{3}\sum_{\mathbf{i}}\left(a_{\mathbf{i}+\mathbf{a}_{x}}^{\dagger}a_{\mathbf{i}+\mathbf{a}_{x}}+a_{\mathbf{i}+\mathbf{a}_{y}}^{\dagger}a_{\mathbf{i}+\mathbf{a}_{y}}+a_{\mathbf{i}+\mathbf{a}_{z}}^{\dagger}a_{\mathbf{i}+\mathbf{a}_{z}}+a_{\mathbf{i}}^{\dagger}a_{\mathbf{i}}\right).
\end{equation}
Next, we move again to the momentum space by means of the same Fourier
transform
\begin{align}
\delta H & =-NKS^{4}+4KS^{3}\sum_{\mathbf{k}}a_{\mathbf{k}}^{\dagger}a_{\mathbf{k}}.
\end{align}
Such an expression for $\delta H$ leads to the following one for
$H'$:
\begin{align}
H' & =-N\left(K_{z}+KS^{2}\right)S^{2}+\sum_{\mathbf{k}}\left[A\left(\mathbf{k}\right)a_{\mathbf{k}}^{\dagger}a_{\mathbf{k}}-\dfrac{1}{2}B\left(\mathbf{k}\right)\left(a_{\mathbf{k}}^{\dagger}a_{\mathbf{-k}}^{\dagger}+a_{\mathbf{k}}a_{-\mathbf{k}}\right)\right]
\end{align}
where $A\left(\mathbf{k}\right)=A_{0}\left(\mathbf{k}\right)+4KS^{3}=S\left(2K_{z}+4KS^{2}-K_{x}\cos\mathbf{k}_{\mathbf{a}_{x}}-K_{y}\cos\mathbf{k}_{\mathbf{a}_{y}}\right)$
and $B\left(\mathbf{k}\right)=B_{0}\left(\mathbf{k}\right)=S\left(K_{x}\cos\mathbf{k}_{\mathbf{a}_{x}}-K_{y}\cos\mathbf{k}_{\mathbf{a}_{y}}\right)$.
Accordingly, the new spin-wave dispersion $\omega'\left(\mathbf{k}\right)$
is
\begin{align}
\omega'\left(\mathbf{k}\right) & =\sqrt{A^{2}\left(\mathbf{k}\right)-B^{2}\left(\mathbf{k}\right)}\nonumber \\
 & =2S\sqrt{\left(K_{z}+2KS^{2}-K_{x}\cos\mathbf{k}_{\mathbf{a}_{x}}\right)\left(K_{z}+2KS^{2}-K_{y}\cos\mathbf{k}_{\mathbf{a}_{y}}\right)}\label{wk}
\end{align}
Comparing the new dispersion $\omega'\left(\mathbf{k}\right)$ to
the previous one $\omega\left(\mathbf{k}\right)$, Eq.~\eqref{wk0},
it is evident that the net effect of the $\delta H$ term in the effective
Hamiltonian $H'$ is the introduction of a rigid (not momentum dependent)
shift of $2KS^{2}$ to the Hamiltonian parameter $K_{z}$. Let us
analyze in detail the new dispersion $\omega'\left(\mathbf{k}\right)$
in the next section and comment briefly on the effects of such a shift.

\begin{figure}[!t]
\noindent \begin{centering}
\includegraphics[width=0.95\textwidth]{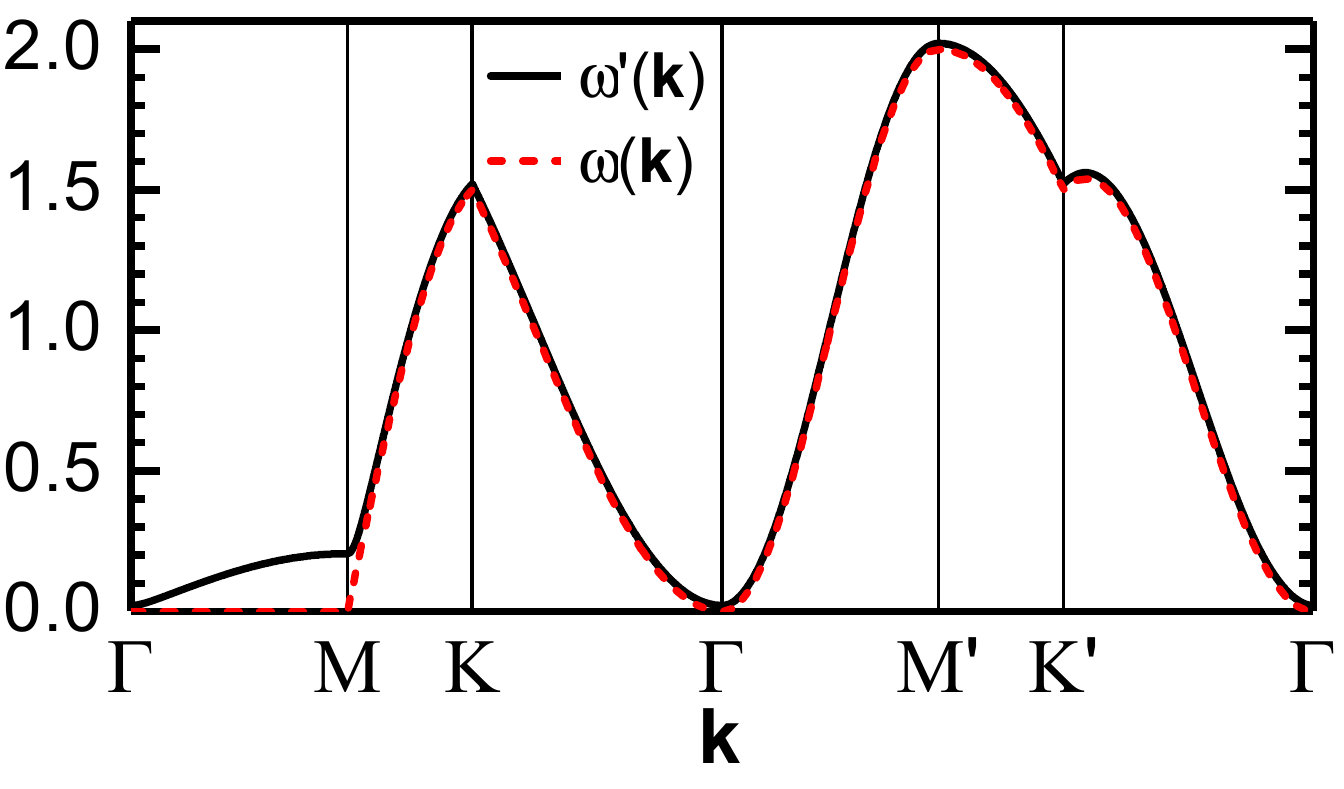}
\par\end{centering}
\caption{Spin-wave excitation spectra $\omega'\left(\mathbf{k}\right)$ and
$\omega\left(\mathbf{k}\right)$ for $K_{\gamma}=1$ and $S=1/2$
along the principal directions of the first Brillouin zone ($\Gamma\rightarrow M\rightarrow K\rightarrow\Gamma\rightarrow M'\rightarrow K'\rightarrow\Gamma$),
see dashed red and dotted blue paths in Fig.~\ref{fig1} (right).\label{fig3}}
\end{figure}

\section{Results}

In Fig.~\ref{fig3}, we report the spin-wave excitation spectra $\omega'\left(\mathbf{k}\right)$,
Eq.~\eqref{wk}, and $\omega\left(\mathbf{k}\right)$, Eq.~\eqref{wk0},
for $K_{\gamma}=1$ and $S=1/2$ along the principal directions of
the first Brillouin zone ($\Gamma\rightarrow M\rightarrow K\rightarrow\Gamma\rightarrow M'\rightarrow K'\rightarrow\Gamma$),
see dashed red and dotted blue paths in Fig.~\ref{fig1} (right).
As already derived and discussed in Ref.~\citep{Jackeli_15}, the
spin-wave excitation spectrum $\omega\left(\mathbf{k}\right)$ is
gapless along the nodal line $\Gamma\rightarrow M$ and all other
equivalent-by-symmetry lines in momentum space because of the sub-extensive
degeneracy of the classical manifold \citep{Jackeli_15}. Then, it
is very remarkable to see that the effective coupling $\delta H$
between pairs of spins belonging to next-nearest-neighbor chains,
emerging from a careful treatment of the quantum fluctuations \citep{Jackeli_15},
is capable to open up a spin gap along such nodal lines. In particular,
for $K_{\gamma}=1$, we have a gap of $\frac{1}{6}S^{3}$ at the bottom
of the band (the $\Gamma$ point) and a gap of $\frac{1}{6}S^{3}\sqrt{1+24/S^{2}}$
at the $M$ point. In fact, the previous nodal line $\Gamma\rightarrow M$
acquires a well defined dispersion because of the interplay between
the two terms under the square root in Eq.~\eqref{wk}: the first
of them, $K_{z}+2KS^{2}-K_{x}\cos\mathbf{k}_{\mathbf{a}_{x}}$, is
no longer identically zero along $\Gamma\rightarrow M$, and this
allows the second one, $K_{z}+2KS^{2}-K_{y}\cos\mathbf{k}_{\mathbf{a}_{y}}$,
to provide a dispersion. The rest of the spin-wave excitation spectrum,
that is along all other explored lines, is minimally affected as expected.
It is worth reminding that $\delta H$ cannot fully lift the degeneracy
of the ground state as the two sublattices formed by nearest-neighbor
chains remain decoupled because of a hidden symmetry of the model
\citep{Jackeli_15}.

\section{Conclusions }

In this short paper, we have analyzed the effects on the spin-wave
excitation spectrum of the triangular analog of the Kitaev model \citep{Bec15,Jackeli_15,Li15}
of an effective coupling between pairs of spins belonging to next-nearest-neighbor
chains emerged by a careful treatment of the quantum fluctuations
within the linked-cluster expansion \citep{Gel00} at the $4^{th}$
order \citep{Jackeli_15}: the $K_{x}$ and $K_{y}$ terms induce
quantum corrections to the ferromagnetic classical ground state. In
absence of such a coupling, the spin-wave excitation spectrum presents
nodal lines along the $\Gamma\rightarrow M$ line and all other equivalent-by-symmetry
lines in momentum space \citep{Jackeli_15}. It is really remarkable
that this effective coupling manages to open up a spin gap in the
spectrum, that becomes fully gapped, and induces a well defined dispersion
along the $\Gamma\rightarrow M$ line (partially removing the degeneracy
in the system), while the rest of the spectrum is minimally affected
as expected.



\begin{thebibliography}{10}
\expandafter\ifx\csname url\endcsname\relax
  \def\url#1{\texttt{#1}}\fi
\expandafter\ifx\csname urlprefix\endcsname\relax\def\urlprefix{URL }\fi
\expandafter\ifx\csname href\endcsname\relax
  \def\href#1#2{#2} \def\path#1{#1}\fi

\bibitem{Bal10}
L.~Balents, Nature 464 (2010) 199.

\bibitem{Vil80}
J.~Villain, R.~Bidaux, J.-P. Carton, R.~Conte, J. Phys. France 41 (1980) 1263.

\bibitem{She82}
E.~F. Shender, Sov. Phys. JETP 56 (1982) 178.

\bibitem{Sav12}
L.~Savary, K.~A. Ross, B.~D. Gaulin, J.~P.~C. Ruff, L.~Balents, Phys. Rev.
  Lett. 109 (2012) 167201.

\bibitem{Jac09}
G.~Jackeli, G.~Khaliullin, Phys. Rev. Lett. 102 (2009) 017205.

\bibitem{Kit06}
A.~Kitaev, Annals of Physics 321 (2006) 2.

\bibitem{Cha10}
J.~Chaloupka, G.~Jackeli, G.~Khaliullin, Phys. Rev. Lett. 105 (2010) 027204.

\bibitem{Jia11}
H.-C. Jiang, Z.-C. Gu, X.-L. Qi, S.~Trebst, Phys. Rev. B 83 (2011) 245104.

\bibitem{Reu11}
J.~Reuther, R.~Thomale, S.~Trebst, Phys. Rev. B 84 (2011) 100406.

\bibitem{Pri12}
C.~C. Price, N.~B. Perkins, Phys. Rev. Lett. 109 (2012) 187201.

\bibitem{Cha13}
J.~Chaloupka, G.~Jackeli, G.~Khaliullin, Phys. Rev. Lett. 110 (2013) 097204.

\bibitem{Sin10}
Y.~Singh, P.~Gegenwart, Phys. Rev. B 82 (2010) 064412.

\bibitem{Liu11}
X.~Liu, T.~Berlijn, W.-G. Yin, W.~Ku, A.~Tsvelik, Y.-J. Kim, H.~Gretarsson,
  Y.~Singh, P.~Gegenwart, J.~P. Hill, Phys. Rev. B 83 (2011) 220403.

\bibitem{Cho12}
S.~K. Choi, R.~Coldea, A.~N. Kolmogorov, T.~Lancaster, I.~I. Mazin, S.~J.
  Blundell, P.~G. Radaelli, Y.~Singh, P.~Gegenwart, K.~R. Choi, S.-W. Cheong,
  P.~J. Baker, C.~Stock, J.~Taylor, Phys. Rev. Lett. 108 (2012) 127204.

\bibitem{Sin12}
Y.~Singh, S.~Manni, J.~Reuther, T.~Berlijn, R.~Thomale, W.~Ku, S.~Trebst,
  P.~Gegenwart, Phys. Rev. Lett. 108 (2012) 127203.

\bibitem{Ye12}
F.~Ye, S.~Chi, H.~Cao, B.~C. Chakoumakos, J.~A. Fernandez-Baca, R.~Custelcean,
  T.~F. Qi, O.~B. Korneta, G.~Cao, Phys. Rev. B 85 (2012) 180403.

\bibitem{Gre13}
H.~Gretarsson, J.~P. Clancy, Y.~Singh, P.~Gegenwart, J.~P. Hill, J.~Kim, M.~H.
  Upton, A.~H. Said, D.~Casa, T.~Gog, Y.-J. Kim, Phys. Rev. B 87 (2013) 220407.

\bibitem{Kha05}
G.~Khaliullin, Prog. Theor. Phys. Suppl. 160 (2005) 155.

\bibitem{Rou16}
I.~Rousochatzakis, U.~K. R\"ossler, J.~van~den Brink, M.~Daghofer, Phys. Rev. B
  93 (2016) 104417.

\bibitem{Bec15}
M.~{Becker}, M.~{Hermanns}, B.~{Bauer}, M.~{Garst}, S.~{Trebst}, Phys. Rev. B
  91 (2015) 155135.

\bibitem{Li15}
K.~Li, S.-L. Yu, J.-X. Li, New Journal of Physics 17~(4) (2015) 043032.

\bibitem{Gel00}
M.~P. Gelfand, R.~R.~P. Singh, Adv. Phys. 49 (2000) 93.

\bibitem{Jackeli_15}
G.~Jackeli, A.~Avella, Phys. Rev. B 92 (2015) 184416.

\end{thebibliography}

\end{document}